\documentclass[preprint,superscriptaddress,amsmath,amssymb,aps,pra]{revtex4-2}

\usepackage{graphicx}
\usepackage{dcolumn}
\usepackage{bm}
\usepackage{gensymb}
\usepackage{url}
\usepackage{multirow}

\usepackage{xcolor}

\begin{document}

\title{A tutorial on the conservation of momentum in \\ photonic time-varying media}

\author{Angel Ortega-Gomez}
\affiliation{%
 Department of Electrical, Electronic and Communications Engineering, Institute of Smart Cities (ISC), Public University of Navarre (UPNA), 31006 Pamplona, Spain
}%

\author{Michaël Lobet}
\affiliation{%
John A. Paulson School of Engineering and Applied Sciences, Harvard University, 9 Oxford Street, Cambridge, MA 02138, USA
}%
\affiliation{%
Department of Physics and Namur Institute of Structured Materials, University of Namur, Rue de Bruxelles 61, 5000 Namur, Belgium
}%

\author{J. Enrique Vázquez-Lozano}
\affiliation{%
 Department of Electrical, Electronic and Communications Engineering, Institute of Smart Cities (ISC), Public University of Navarre (UPNA), 31006 Pamplona, Spain
}%

\author{Iñigo Liberal}
\thanks{Corresponding author: inigo.liberal@unavarra.es}%
\affiliation{%
 Department of Electrical, Electronic and Communications Engineering, Institute of Smart Cities (ISC), Public University of Navarre (UPNA), 31006 Pamplona, Spain
}%

\begin{abstract}
Time-varying media break temporal symmetries while preserving spatial symmetries intact. Thus, it represents an excellent conceptual framework to investigate the fundamental implications of Noether's theorem for the electromagnetic field. At the same time, addressing momentum conservation in time-varying media sheds light on the Abraham-Minkowski debate, where two opposing forms of the electromagnetic field momentum are defended. Here, we present a tutorial review on the conservation of momentum in time-varying media. We demonstrate that the Minkowski momentum is a conserved quantity with three independent approaches of increasing complexity: (i) via the application of the boundary conditions for Maxwell equations at a temporal boundary, (ii) testing for constants of motion and deriving conservation laws, and (iii) applying temporal and spatial translations within the framework of the Lagrangian theory of the electromagnetic field. Each approach provides a different and complementary insight into the problem.  
\end{abstract}

\maketitle

\section{Introduction}

Time-varying media are revolutionizing the fields of optics and nanophotonics by harnessing time as an additional resource for controlling light-matter interactions \cite{Caloz2019spacetime,Galiffi2022photonics,Yin2022,Engheta2021metamaterials}. Dynamically modulating matter offers new possibilities for the manipulation of electromagnetic fields including compact and low-energy nonreciprocal devices \cite{Sounas2017non}, inverse prism and temporal aiming effects \cite{akbarzadeh2018inverse,Pacheco2020temporal}, overcoming bandwidth bounds in impedance matching \cite{shlivinski2018beyond}, energy accumulation without a theoretical limit \cite{Mirmoosa2019time}, quantum state frequency shifting \cite{Liberal2022quantum}, and ultra-fast switching without thermal noise amplification \cite{Liberal2022quantum}, to name a few. Time-varying media also empower new amplification \cite{Pendry2021gain} and photon generation mechanisms, such as directional vacuum amplification effects \cite{Vazquez2022shaping}, amplified light emission from quantum emitters \cite{Lyubarov2022amplified} and free electrons \cite{Dikopoltsev2022light}, as well as incandescent sources not constrained within the black-body spectrum \cite{Vazquez2022incandescent}. 

Because a homogeneous time-varying medium is invariant under spatial translations (see Fig.1), it is usually argued that time-varying media preserves the momentum of the electromagnetic field \cite{Caloz2019spacetime,Galiffi2022photonics,Yin2022,Engheta2021metamaterials,Morgenthaler1958velocity,Solis2021time,Koutserimpas2020electromagnetic}. This intuition stems from Noether's theorem \cite{Banados2016short,Kosmann2011noether,Cohen1997photons,Sakurai2014}, which more generally states that symmetries of the action of a physical system have an associated conserved quantity. However, a direct connection between invariance under spatial translations and momentum conservation in time-varying media is not specified. In addition, the notion of the momentum of the electromagnetic field is quite subtle. In fact, according to the Abraham-Minkowski debate \cite{Brevik1979experiments,Pfeifer2007colloquium,Kemp2011,Milonni2005recoil,Mansuripur2004radiation}, there is more than one definition for the momentum of the electromagnetic field. On the one hand, one can define the Abraham momentum,

\begin{equation}
\mathbf{P}_{A}\left(t\right)
=\int d^{3}r\,\,\,\mathbf{p}_{A}\left(\mathbf{r},t\right)
=\mu_0\varepsilon_0\int d^{3}r\,\,\,\mathbf{E}\left(\mathbf{r},t\right)\times\mathbf{H}\left(\mathbf{r},t\right)
\label{eq:P_A}
\end{equation}
 
\noindent where we have also defined the Abraham momentum density, which is proportional to the Poynting vector field,
$\mathbf{p}_{A}=\mu_0\varepsilon_0\mathbf{S}$. On the other hand, the Minkowski momentum reads as
\begin{equation}
\mathbf{P}_{M}\left(t\right)
=\int d^{3}r\,\,\,\mathbf{p}_{M}\left(\mathbf{r},t\right)
=\int d^{3}r\,\,\,\mathbf{D}\left(\mathbf{r},t\right)\times\mathbf{B}\left(\mathbf{r},t\right)
\label{eq:P_M}
\end{equation}

A common simplification of those definitions for a plane wave in non-dispersive media is $p_A=\hbar\omega/nc$ and $p_M=n \hbar\omega/c$, which highlights the role of the refractive index $n$. Interestingly, as pointed out by Leonhardt \cite{Leonhardt2006momentum} one should call for the Minkowski momentum whenever the wave aspects dominate, for example, in experiments involving momentum recoil \cite{Campbell2005photon,Milonni2005recoil}, while  the Abraham momentum appears when the particle aspects are probed \cite{Brevik1979experiments}.

A resolution of the debate was offered among others by Barnett \cite{Barnett2010resolution,Barnett2010enigma}. It is suggested that the Abraham momentum is the kinetic momentum of the electromagnetic field, associated with energy transport. The Minkowski momentum is, however, the canonical momentum of the electromagnetic field, being the generator of spatial translations. Nevertheless, certain aspects of the momentum of the electromagnetic field are still under question \cite{Silveirinha2017reexamination}. Moreover, the avenue of near-zero-index (NZI) media exacerbates the differences between the forms of the momentum \cite{Lobet2022momentum,Liberal2017near,kinsey2022developing} giving rise to zero Minkowski momentum but nonzero Abraham momentum inside epsilon-and-mu-near-zero (EMNZ) media where both permittivity and permeability approach zero.

Since time-varying media preserve spatial symmetries while breaking temporal symmetries, it represents an excellent conceptual playground to illuminate the Abraham-Minkowski debate. Following the interpretation offered by  Barnett  \cite{Barnett2010resolution}, it should be expected that the Minkowski momentum - related to spatial translations - is a conserved quantity, while the Abraham momentum - related to energy transport - is not. This work aims to provide a tutorial review of different aspects on the conservation of the momentum of the electromagnetic field in time-varying media. We address three independent derivations showing that only the Minkowski momentum is a conserved quantity in time-varying media based on: (i) boundary conditions on Maxwell equations, (ii) directly evaluating constants of motion and deriving conservation laws, and (iii) inducing spatial translations to the Lagrangian of the electromagnetic field. Each approach provides a different physical insight into the problem. 

\begin{figure*}
\includegraphics[width=4.0in]{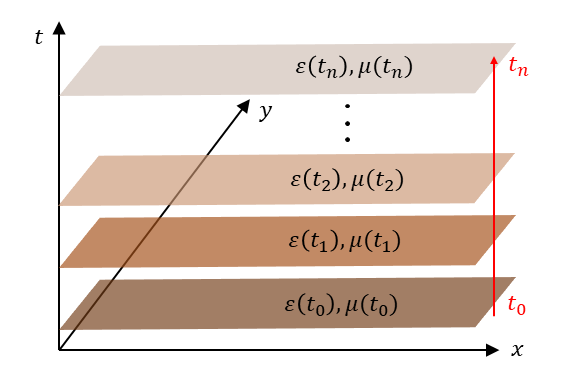}
\caption{Schematic depiction of time-varying media, in which both permittivity $\varepsilon\left(t\right)$ and permeability $\mu\left(t\right)$ change with time. Thus, the systems is invariant with respect to spatial translations, but is not invariant with respect to temporal translations.}
\label{fig:Time_varying}
\end{figure*}

\section{Momentum conservation from inspecting Maxwell equations at a temporal boundary}

Our starting point is Maxwell curl equations in time-varying media, which, in the absence of charges and currents, can be written as follows
\begin{equation}
\nabla\times\mathbf{E}\left(\mathbf{r},t\right)=-\partial_{t}\mathbf{B}\left(\mathbf{r},t\right)
\label{eq:curl_E}
\end{equation}
\begin{equation}
\nabla\times\mathbf{H}\left(\mathbf{r},t\right)=\partial_{t}\mathbf{D}\left(\mathbf{r},t\right)
\label{eq:curl_H}
\end{equation}

For the sake of simplicity, we assume homogeneous and instantaneous time-varying media, with constitutive relations
\begin{equation}
\mathbf{D}\left(\mathbf{r},t\right)=\varepsilon\left(t\right)\mathbf{E}\left(\mathbf{r},t\right)
\label{eq:Constitutive_D}
\end{equation}

\begin{equation}
\mathbf{B}\left(\mathbf{r},t\right)=\mu\left(t\right)\mathbf{H}\left(\mathbf{r},t\right)
\label{eq:Constitutive_H}
\end{equation}

A more complete description of time-varying media would include the impact of dispersion and loss \cite{Solis2021time,Gratus2021}. However, a system with dissipation does not necessarily conserve quantities even in the presence of symmetries. In addition, the assumption of instantaneous media is  widespread in the field of temporal metamaterials \cite{Galiffi2022photonics}. 

Integrating Maxwell equations (\ref{eq:curl_E})-(\ref{eq:curl_H}) accross a temporal boundary taking place at $t_0$, where material parameters suddenly change from $\varepsilon(t_0^{-})$, $\mu(t_0^{-})$ to $\varepsilon(t_0^{+})$, $\mu(t_0^{+})$, gives

\begin{equation}
\int_{t_{0}^{-}}^{t_{0}^{+}}\,\,dt\,\nabla\times\mathbf{H}\left(\mathbf{r},t\right)=
\int_{t_{0}^{-}}^{t_{0}^{+}}dt\,\,\partial_{t}\mathbf{D}\left(\mathbf{r},t\right)=\mathbf{D}\left(\mathbf{r},t_{0}^{+}\right)-\mathbf{D}\left(\mathbf{r},t_{0}^{-}\right)
\end{equation}

\begin{equation}
-\int_{t_{0}^{-}}^{t_{0}^{+}}\,\,dt\,\nabla\times\mathbf{E}\left(\mathbf{r},t\right)=
\int_{t_{0}^{-}}^{t_{0}^{+}}dt\,\,\partial_{t}\mathbf{B}\left(\mathbf{r},t\right)=\mathbf{B}\left(\mathbf{r},t_{0}^{+}\right)-\mathbf{B}\left(\mathbf{r},t_{0}^{-}\right)
\end{equation}

Therefore, we find that  $\mathbf{D}\left(\mathbf{r},t\right)$ and $\mathbf{B}\left(\mathbf{r},t\right)$  must be continuous accross changes of the constitutive parameters, for finite $\mathbf{E}\left(\mathbf{r},t\right)$ and $\mathbf{H}\left(\mathbf{r},t\right)$ fields. This property is well-known since early works on time-varying media \cite{Morgenthaler1958velocity}. As a consequence, this reasoning confirms that the Minkowski momentum – uniquely defined as a function of $\mathbf{D}$ and $\mathbf{B}$ fields via Eq.\,(\ref{eq:P_M}) - is a continuous quantity across a temporal boundary, suggesting that is should be a  conserved quantity in time-varying media. However, this approach fails at providing any insight on the associated conservation law and/or how it can be related to invariance under spatial translations. Moreover, it does not clarify the (non) conservation of Abraham momentum.

\section{Constants of motion and conservation laws}

In this section, we address the conservation of momentum in time-varying media by direcly testing if a given quantity is a constant of motion. To this end, one can take the time derivative of the quantity under question and check if it is zero, in which case it shall be a constant of motion/conserved quantity. Before addressing the momentum, it is instructive to analyze the energy of the electromagnetic field, which in time-varying media can be written as
\begin{equation}
U\left(t\right)=\int d^{3}r\,\,\,u\left(\mathbf{r},t\right)
\label{eq:H}
\end{equation}

\noindent with energy density
\begin{equation}
u\left(\mathbf{r},t\right)=\frac{1}{2}\left[\varepsilon\left(t\right)E^{2}\left(\mathbf{r},t\right)+\mu\left(t\right)H^{2}\left(\mathbf{r},t\right)\right]\label{eq:h}
\end{equation}

Taking the time derivative of the energy and, substituting Maxwell equations (\ref{eq:curl_E})-(\ref{eq:curl_H}), leads to the following expression 
\begin{equation}
\frac{dU}{dt}=-\frac{1}{\mu_{0}\varepsilon_{0}}\int d\mathbf{S}\cdot\mathbf{p}_{A}-\frac{1}{2}\int d^{3}r\,\left[\frac{d\varepsilon\left(t\right)}{dt}E^{2}\left(\mathbf{r},t\right)+\frac{d\mu\left(t\right)}{dt}H^{2}\left(\mathbf{r},t\right)\right]
\label{eq:dtH}
\end{equation}

On the one hand, the first term in the r.h.s. of (\ref{eq:dtH}) is a surface term proportional to the $\mathbf{E}$ and $\mathbf{H}$ fields. This term physically means that the change of energy over time is partly due to energy either leaking out or coming into the system. It can be seen as a flux of either outgoing or incoming Poynting vector field, hence setting down a link with $\mathbf{P}_{A}$ (Eq.\,(\ref{eq:P_A})). It confirms the role of the Abraham momentum as the kinetic momentum, associated with energy transport. If the volume is large enough to capture the entirety of the $\mathbf{E}$ and $\mathbf{H}$ fields within the time interval of interest, its contribution vanishes. On the other hand, the second term in the r.h.s. of (\ref{eq:dtH}) is a volume integral directly linked to the time modulation of the permittivity and permeability, which results in a change of the energy of the system. It represents the energy that must be pumped into or retracted from the system in order to realize the time modulation of the material parameters. In other words, the time variation of the material parameters act as sources or sinks of electromagnetic energy. By contrast, Eq.\,(\ref{eq:dtH}) shows that for a medium with static material properties $dU/dt=0$ and energy would be a conserved quantity.

Eq.\,(\ref{eq:dtH}) can also be casted as a local conservation law as a function of the energy and momentum densities
\begin{equation}
\frac{du\left(\mathbf{r},t\right)}{dt}+\frac{1}{\mu_{0}\varepsilon_{0}}\nabla\cdot\mathbf{p}_{A}\left(\mathbf{r},t\right)=-\frac{1}{2}\left[\frac{d\varepsilon\left(t\right)}{dt}E^{2}\left(\mathbf{r},t\right)+\frac{d\mu\left(t\right)}{dt}H^{2}\left(\mathbf{r},t\right)\right]
\label{eq:dtH_local}
\end{equation}

\noindent where we clearly identify the source/sink at the r.h.s..

Let us now tackle the conservation of Minkowski momentum and examine the time variation of Abraham momentum. By introducing Maxwell equations and applying a few vector calculus identities, it can be found that the time derivative of the Minkowski momentum is given by
\[
\frac{d\mathbf{P}_{M}\left(t\right)}{dt}=\varepsilon\left(t\right)\,\left[\sum_{p=x,y,z}\mathbf{u}_{p}\int d\mathbf{S}\cdot\left(E_{p}\mathbf{E}\right)-\frac{1}{2}\,\int d\mathbf{S}\left(\mathbf{E}\cdot\mathbf{E}\right)\right]
\]
\begin{equation}
+\mu\left(t\right)\,\left[\sum_{p=x,y,z}\mathbf{u}_{p}\int d\mathbf{S}\cdot\left(H_{p}\mathbf{H}\right)-\frac{1}{2}\,\int d\mathbf{S}\left(\mathbf{H}\cdot\mathbf{H}\right)\right]
\label{eq:dP_M}
\end{equation}

By doing so, we find that the time derivative of the Minkowski momentum reduces to surface terms. Once again, if the volume of integration is taken large enough so that all the  $\mathbf{E}$ and $\mathbf{H}$ fields are confined within its interior, all surface terms vanish. In other words, $d\mathbf{P}_{M}\left(t\right)/dt=0$, proving that the Minkowski momentum is a constant of motion as expected. It is also instructive to note that the above equation can be written in a differential form as a conservation law for the momentum density:
\begin{equation}
\frac{d\mathbf{p}_{M}\left(t\right)}{dt}=\nabla\cdot\overline{\overline{\mathbf{T}}}_{M}\left(\mathbf{r},t\right)
\label{eq:p_M_Conservation_law}
\end{equation}

\noindent where we define the Minkowski stress tensor for time-varying media as
\begin{equation}
\overline{\overline{\mathbf{T}}}_{M}\left(\mathbf{r},t\right)=\varepsilon\left(t\right)\left(\mathbf{E}\otimes\mathbf{E}-\frac{1}{2}\overline{\overline{\mathbf{I}}}\,\left(\mathbf{E}\cdot\mathbf{E}\right)\right)+\mu\left(t\right)\left(\mathbf{H}\otimes\mathbf{H}-\frac{1}{2}\overline{\overline{\mathbf{I}}}\left(\mathbf{H}\cdot\mathbf{H}\right)\right)
\label{eq:T_M}
\end{equation}

\noindent with $\overline{\overline{\mathbf{I}}}$ being the identity dyadic. Conservation laws in the form of (\ref{eq:p_M_Conservation_law})
can be found scattered in the literature, for example, in the appendix of \cite{Koutserimpas2020electromagnetic}.

Proceeding similarly with the Abraham momentum reveals that in general it is not a conserved quantity:
\[
\frac{d\mathbf{P}_{A}}{dt}=-\left(\frac{1}{\varepsilon\left(t\right)}\frac{d\varepsilon\left(t\right)}{dt}+\frac{1}{\mu\left(t\right)}\frac{d\mu\left(t\right)}{dt}\right)\mathbf{P}_{A}\left(t\right)
\]
\[
-\frac{\varepsilon_{0}\mu_{0}}{\mu\left(t\right)}\left[\frac{1}{2}\,\int d\mathbf{S}\left(\mathbf{E}\cdot\mathbf{E}\right)-\sum_{p=x,y,z}\mathbf{u}_{p}\int d\mathbf{S}\cdot\left(E_{p}\mathbf{E}\right)\right]
\]
\begin{equation}
-\frac{\varepsilon_{0}\mu_{0}}{\varepsilon\left(t\right)}\,\left[\frac{1}{2}\,\int d\mathbf{S}\left(\mathbf{H}\cdot\mathbf{H}\right)-\sum_{p=x,y,z}\mathbf{u}_{p}\int d\mathbf{S}\cdot\left(H_{p}\mathbf{H}\right)\right]
\label{eq:dtP_A}
\end{equation}

Here again, the second and third terms are surface terms that would vanish for a sufficiently large volume. However, the first term illustrates that the Abraham momentum does change in time, following the change in the permittivity and permeability of the medium. Equation (\ref{eq:dtP_A}) can also be compactly written as a local conservation law for the momentum density
\begin{equation}
\frac{d\mathbf{p}_{A}}{dt}=\nabla\cdot\overline{\overline{\mathbf{T}}}_{A}-\left(\frac{1}{\varepsilon\left(t\right)}\frac{d\varepsilon\left(t\right)}{dt}+\frac{1}{\mu\left(t\right)}\frac{d\mu\left(t\right)}{dt}\right)\mathbf{p}_{A}\left(\mathbf{r},t\right)
\end{equation}

\noindent where we define the Abraham stress tensor in time-varying media, related to the Minkowski stress tensor as follows  
\begin{equation}
\overline{\overline{\mathbf{T}}}_{A}=\frac{\varepsilon_{0}\mu_{0}}{\mu\left(t\right)\varepsilon\left(t\right)}\,\overline{\overline{\mathbf{T}}}_{M}\left(\mathbf{r},t\right)
\end{equation}

In conclusion, testing for constants of motions provides an independent confirmation that the Minkowski momentum is indeed a conserved quantity in time-varying media. In addition, it provides insight in the form of the conservation law that supports its invariance. Furthermore, it shows that the Abraham momentum is not a constant of motion in close connection to energy considerations, and re-emphasizes its role as the kinetic momentum of the electromagnetic field. Nevertheless, writing the conservation law does not clarify the role of the invariance of the system under spatial translations in the conservation of momentum.

\section{Momentum conservation as a consequence of invariance under spatial translations: A Lagrangian approach}

In this section we address the conservation of momentum in time-varying media from the perspective of the Lagrangian formalism for electromagnetic fields. Using the Lagrangian formalism adds an extra layer of complexity, but allows to unequivocally identify momentum conservation as a fundamental consequence of the invariance of time-varying media under spatial translations. We note that most works identifying the Minkowski momentum as the generator of spatial translations do it from a quantum description of the electromagnetic field, where the Minkowski momentum appears as an operator \cite{Barnett2010resolution}. However, it is important to understand that momentum conservation as a consequence of invariance under spatial translations is also a classical effect. Therefore, we keep here a classical Lagrangian description of the electromagnetic fields, without introducing the quantization of the electromagnetic field.

In the following, we first review the Lagrangian description of electromagnetic fields extended to time-varying media. Then, we derive a form of Noether's theorem in our formalism and we finally show the quantities associated with temporal and spatial translations for time-varying media.

\subsection{Lagrangian description of the electromagnetic field}

An in-depth review of the Lagrangian theory of the electromagnetic field can be found in Cohen-Tannoudji's book \cite{Cohen1997photons}. Here we review it and extend it to time-varying media. From the perspective of Lagrangian theory, Maxwell equations are equations of motion that can be derived from the principle of least (or stationary) action. This principle states that true path of motion corresponds to a stationary point of the action. By motion we refer to the values that the dynamical variables have in a given interval of time, which, when position is a dynamical variable, aligns with the common notion of  motion. The action is defined as the integral of the Lagrangian between two instants of time $t_{1}$ and $t_{2}$:
\begin{equation}
S\left(t_{1},t_{2}\right)=\int_{t_{1}}^{t_{2}}dt\,L\left(t\right)
\end{equation}

\noindent with the Lagrangian
\begin{equation}
L\left(t\right)=\int d^{3}r\,\mathcal{L}\left(\mathbf{r},t\right)
\end{equation}

\noindent and the Lagrangian density
\begin{equation}
\mathcal{L}\left(\mathbf{r},t\right)=\frac{1}{2}\int d^{3}r\,\left[\varepsilon\left(t\right)\mathbf{E}\left(\mathbf{r},t\right)\cdot\mathbf{E}\left(\mathbf{r},t\right)-\mu\left(t\right)\mathbf{H}\left(\mathbf{r},t\right)\cdot\mathbf{H}\left(\mathbf{r},t\right)\right]
\end{equation}

The choice of this Lagrangian density is a direct extension from the case with no time modulation. It is justified because Lagrange's equation correctly recovers the equations of motion for the electromagnetic field, as shown below. For the Lagrangian description of the electromagnetic field, it is convenient to work with scalar 
$V\left(\mathbf{r},t\right)$ and vector $\mathbf{A}\left(\mathbf{r},t\right)$ potentials instead of fields. For the sake of simplicity, we work in the Coulomb gauge, for which 
$\nabla\cdot\mathbf{A}\left(\mathbf{r},t\right)=0$. By doing so, the scalar potential is zero in the absence of charges $V\left(\mathbf{r},t\right)=0$, all the fields are transversal, and they can be simply written as a function of the vector potential
\begin{equation}
\mathbf{D}\left(\mathbf{r},t\right)=-\varepsilon\left(t\right)\partial_{t}\mathbf{A}\left(\mathbf{r},t\right)
\end{equation}
\begin{equation}
\mathbf{B}\left(\mathbf{r},t\right)=\nabla\times\mathbf{A}\left(\mathbf{r},t\right)
\end{equation}

Then, Maxwell equations lead to the following wave equation for the components of the vector potential ($p=x,y,z$):
\begin{equation}
\nabla^{2}A_{p}\left(\mathbf{r},t\right)-\mu\left(t\right)\partial_{t}\left\{ \varepsilon\left(t\right)\partial_{t}A_{p}\left(\mathbf{r},t\right)\right\} =0
\label{eq:Wave_eq_Aj}
\end{equation}

Due to field transversality, the Minkowski momentum can be compactly written as
\begin{equation}
\mathbf{P}_{M}=-\varepsilon\left(t\right)\sum_{p}\int d^{3}r\,\partial_{t}A_{p}\left(\mathbf{r},t\right)\nabla A_{p}\left(\mathbf{r},t\right)
\label{eq:P_M_A}
\end{equation}

Similarly, the Lagrangian density reduces to
\begin{equation}
\mathcal{L}=\frac{1}{2}\,\sum_{p}\left(\varepsilon\left(t\right)\dot{A}_{p}^{2}\left(\mathbf{r},t\right)-\frac{1}{\mu\left(t\right)}\,\left(\nabla\times\mathbf{A}\left(\mathbf{r},t\right)\right)_{p}^{2}\right)
\label{eq:L_A}
\end{equation}

\noindent where we have used $\dot{A}_{p}$ as a shorter way to write the time derivative. From this description, it lies that the components of the vector potential, $A_{p}$, and its time derivatives, $\dot{A}_{p}$, are the dynamical variables of the system.

Imposing that a true path of motion is a stationary point of the action, for which $\delta S=0$, leads to Lagrange's equations
\begin{equation}
\frac{\partial\mathcal{L}}{\partial A_{p}}-\sum_{q}\,\partial_{q}\left(\frac{\partial\mathcal{L}}{\partial\left(\partial_{q}A_{p}\right)}\right)-\frac{d}{dt}\,\frac{\partial\mathcal{L}}{\partial\dot{A}_{p}}=0
\label{eq:Lagranges_equation}
\end{equation}

\noindent which reduces to the wave equation for $A_{p}$ in (\ref{eq:Wave_eq_Aj}), justifying the direct extension of the Lagrangian to time-varying media. 

With equation (\ref{eq:L_A}), we find that the conjugate momentum of each vector potential component, $A_{p}$, is the negative of the electric displacement field components
\begin{equation}
\varPi_{p}\left(\mathbf{r},t\right)=\frac{\partial\mathcal{L}}{\partial\dot{A}_{p}}=\varepsilon\left(t\right)\,\dot{A}_{p}\left(\mathbf{r},t\right)=-D_{p}\left(\mathbf{r},t\right)
\end{equation}

This point allow us to clarify another ambiguity related to the momentum of the electromagnetic field. For a freely moving particle of mass $m$ with Lagrangian, $L=\sum_{p}\frac{1}{2}\,m\,\dot{r}_{p}^{2}$, the dynamical variables are the position coordinates $r_{p}$, $p=x,y,z$. Thus, their associated conjugate momenta $p_{p}=\partial L/\partial\dot{r}_{p}=m\,\dot{r}_{p}$ correspond to the components of the linear momentum. The latter is also the momentum associated with the spatial translations of the system. However, for the electromagnetic field, position is not a dynamical variable of the system while the vector potential is. For this reason, one has to differentiate between the conjugate momentum and the momentum associated with spatial translations, as clarified below. 

Finally, the Hamiltonian is defined as a function of the conjugate momentum as follows
\begin{equation}
H=\sum_{p}\,\int d^{3}r\,\varPi_{p}\left(\mathbf{r},t\right)\dot{A}_{p}\left(\mathbf{r},t\right)-L
\end{equation}

\noindent which can be found to be fully equivalent to the form of the electromagnetic energy in time-varing media employed in the previous section, and given by Eqs.\,(\ref{eq:H})-(\ref{eq:h}). 

\begin{figure*}
\includegraphics[width=6.25in]{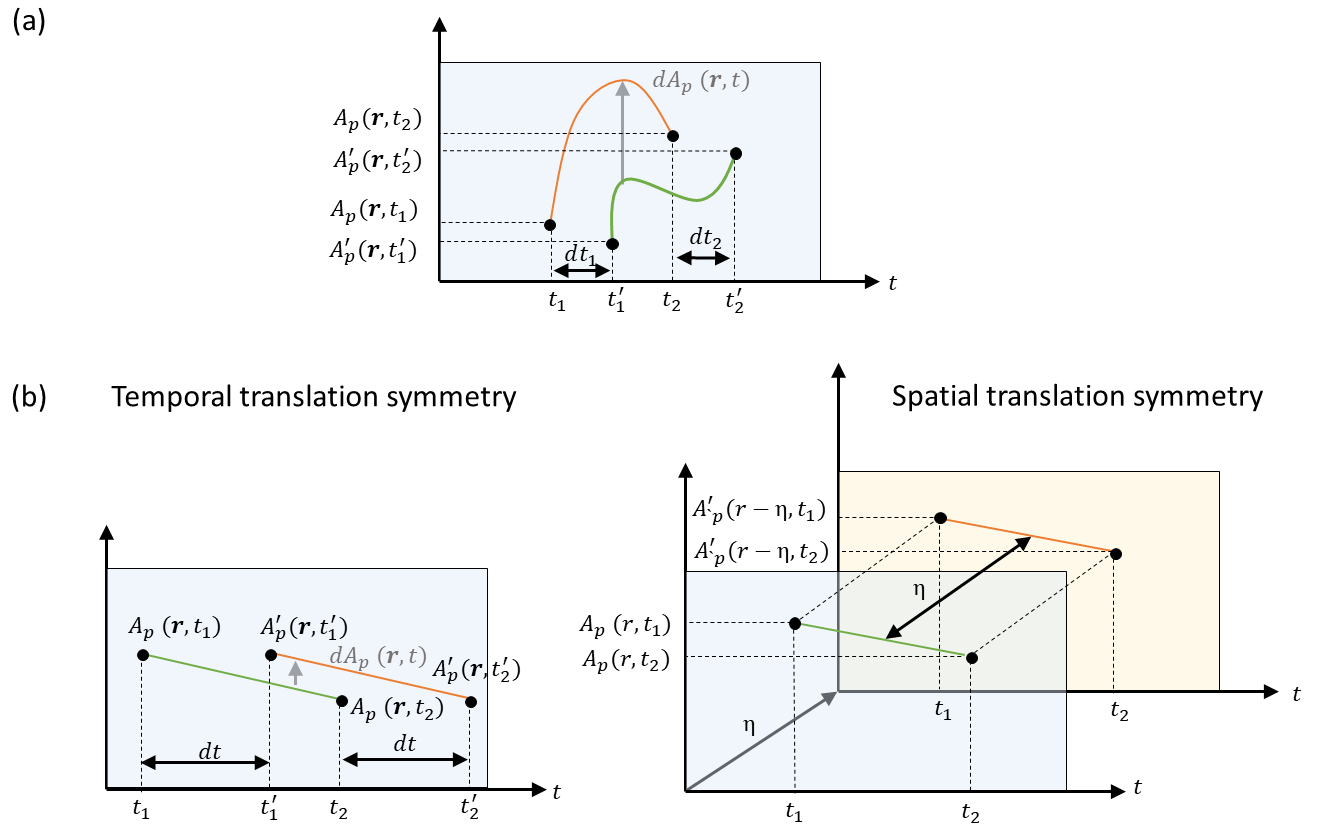}
\caption{(a) Schematic depiction of the motion of a dynamical variable $A_{p}\left(\mathbf{r},t\right)$ between times $t_1$ and $t_2$, and an infinitesimally close motion, described by $A'_{p}\left(\mathbf{r},t\right)$ between times $t_1'$ and $t_2'$. The difference between both motions at time $t$ is given by 
$dA\left(\mathbf{r},t\right)=A'\left(\mathbf{r},t\right)-A\left(\mathbf{r},t\right)$. The difference between the initial and final temporal points is given by 
$dt_1=t_1'-t_1$ and $dt_2=t_2'-t_2$, respectively. (b) Schematic depiction of trajectories for systems with (left) temporal translation symmetry, and (right) spatial translation symmetry.}
\label{fig:Variation}
\end{figure*}

\subsection{Noether's theorem in the Coulomb gauge}

In this section, we cast a form of Noether's theorem which allows us to discern the conserved quantities associated with the continuous symmetries of time-varying media. To this end, we note that any continuous symmetry can be described as an infinitesimal variation of the action. Therefore, as schematically depicted in Fig.\,\ref{fig:Variation}(a), we consider a motion between times $t_{1}$
and $t_{2}$, defined by the dynamical variables $A_{p}\left(\mathbf{r},t\right)$, and an infinitesimally close motion between times $t'_{1}$ and $t'_{2}$,
described by $A'_{p}\left(\mathbf{r},t\right)$. The variation of the dynamical variables at a given point of time is $dA_{p}\left(\mathbf{r},t\right)=A'_{p}\left(\mathbf{r},t\right)-A{}_{p}\left(\mathbf{r},t\right)$, and the variation of the action can be written as
\[
dS=S'-S=\int_{t'_{1}}^{t'_{2}}dt\,L\left(A'_{p}\right)-\int_{t_{1}}^{t_{2}}dt\,L\left(A_{p}\right)
\]
\begin{equation}
=\int_{t_{1}}^{t_{2}}dt\,\left[L\left(A'_{p}\right)-L\left(A_{p}\right)\right]+\int_{t_{2}}^{t'_{2}}dt\,L\left(A'_{p}\right)-\int_{t_{1}}^{t'_{1}}dt\,L\left(A'_{p}\right)
\label{eq:dS}
\end{equation}

To first order, last two terms can be approximated by 
\begin{equation}
\int_{t_{2}}^{t'_{2}}dt\,L\left(A'_{p}\right)=\left.L\left(A_{p}\right)\right|_{t_{2}}dt_{2}
\label{eq:L2_contribution}
\end{equation}

\noindent and the equivalent expression for $t_1$.

Similarly, for two infinitesimally closed motions, the first term is given by
\[
\int_{t_{1}}^{t_{2}}dt\,\left[L\left(A'_{p}\right)-L\left(A_{p}\right)\right]=
\]
\[
=\int_{t_{1}}^{t_{2}}dt\,\sum_{p}\int d^{3}r\,\,\left\{ \frac{\partial\mathcal{L}}{\partial A_{p}\left(\mathbf{r},t\right)}\,dA_{p}\left(\mathbf{r},t\right)+\frac{\partial\mathcal{L}}{\partial\dot{A}_{p}\left(\mathbf{r},t\right)}\,d\dot{A}_{p}\left(\mathbf{r},t\right)
\right.
\]
\begin{equation}
\left.
+\sum_{q}\left(\frac{\partial\mathcal{L}}{\partial\left(\partial_{q}A_{p}\left(\mathbf{r},t\right)\right)}\right)d\partial_{q}A_{p}\left(\mathbf{r},t\right)\right\} 
\end{equation}

Similarly to the derivation of Lagrange's equation, we integrate by parts the second term with respect to time and the third term with respect to $\mathbf{r}$, so the variation of the action reduces to
\[
\int_{t_{1}}^{t_{2}}dt\,\left[L\left(A'_{p}\right)-L\left(A_{p}\right)\right]=
\]
\[
=\int_{t_{1}}^{t_{2}}dt\,\sum_{p}\int d^{3}r\,\,\left\{ \frac{\partial\mathcal{L}}{\partial A_{p}\left(\mathbf{r},t\right)}-\frac{d}{dt}\frac{\partial\mathcal{L}}{\partial\dot{A}_{p}\left(\mathbf{r},t\right)}-\sum_{q}\partial_{q}\left(\frac{\partial\mathcal{L}}{\partial\left(\partial_{q}A_{p}\left(\mathbf{r},t\right)\right)}\right)\right\} dA_{p}\left(\mathbf{r},t\right)
\]
\begin{equation}
+\left.\sum_{p}\int d^{3}r\,\,\frac{\partial\mathcal{L}}{\partial\dot{A}_{p}\left(\mathbf{r},t\right)}\,dA_{p}\left(\mathbf{r},t\right)\right|_{t_{1}}^{t_{2}}
\label{eq:L2_L1}
\end{equation}

Note that, in deriving the above equation we have assumed that the fields vanish at infinity, so that there are no surface contributions. By contrast, the fields do not need to vanish at the initial and final temporal boundaries, leading the contribution from the last term. In addition, the integrand of the first term is a solution to Lagrange's equation (\ref{eq:Lagranges_equation}), which reduces to zero. Thus, by substituting (\ref{eq:L2_contribution})-(\ref{eq:L2_L1}) into (\ref{eq:dS}) we find that the variation of the action is given by: 
\[
dS=\sum_{p}\int d^{3}r\left\{ \left.\frac{\partial\mathcal{L}}{\partial\dot{A}_{p}\left(\mathbf{r},t\right)}\,dA_{p}\left(\mathbf{r},t\right)\right|_{t_{2}}+\left.L\left(A_{p}\right)\right|_{t_{2}}dt_{2}
\right.
\]
\begin{equation}
\left.
-\left.\frac{\partial\mathcal{L}}{\partial\dot{A}_{p}\left(\mathbf{r},t\right)}\,dA_{p}\left(\mathbf{r},t\right)\right|_{t_{1}}-\left.L\left(A_{p}\right)\right|_{t_{1}}dt_1\right\}
\label{eq:dS_final}
\end{equation}

If a system has a continuous symmetry, then the corresponding action remains invariant with respect to infinitesimal displacements, i.e., $dS=0$. In addition, since $dS=0$ must hold for any pair of times $t_{1}$ and $t_{2}$ we find that the term within brackets must be a constant of motion. These relations correspond to Noether's theorem applied to our formulation of the electromagnetic field in time-varying media in the Coulomb Gauge. Given a continuous symmetry, specified by the variation $dA_{p}\left(\mathbf{r},t\right)$ and the boundary condition on the Lagrangian $L\left(A_{p}\right)dt$, one can identify an associated conserved quantity.

\subsection{Temporal and spatial translations}

First, let us assume that the variation is produced by an infinitesimal temporal displacement $dt$, such that $dt_{2}=dt_{1}=dt$ (see Fig.\,\ref{fig:Variation}(b)). If the system is invariant with respect to temporal translations we can write
$A'_{p}\left(\mathbf{r},t\right)=A_{p}\left(\mathbf{r},t-dt\right)\simeq A_{p}\left(\mathbf{r},t\right)-\dot{A}{}_{p}\left(\mathbf{r},t\right)dt$. Then, we have $dA_{p}\left(\mathbf{r},t\right)=-\dot{A}{}_{p}\left(\mathbf{r},t\right)dt$.
Substituting this result in (\ref{eq:dS_final}) and factoring out $dt$ we find that the conserved quantity is
\[
\sum_{p}\int d^{3}r\,\,\left[-\frac{\partial\mathcal{L}}{\partial\dot{A}_{p}\left(\mathbf{r},t\right)}\,\dot{A}_{p}\left(\mathbf{r},t\right)+L\left(A_{p}\right)\right]=
\]
\begin{equation}
=-\sum_{p}\int d^{3}r\,\,\frac{1}{2}\,\sum_{p}\left(\varepsilon\left(t\right)\dot{A}_{p}^{2}\left(\mathbf{r},t\right)+\frac{1}{\mu\left(t\right)}\,\left(\nabla\times\mathbf{A}\left(\mathbf{r},t\right)\right)_{p}^{2}\right)
=-H
\end{equation}

Therefore, it is found that invariance with respect to temporal translations implies that the Hamiltonian must be a conserved quantity. In time-varying media, the system is not invariant under temporal translations, and, consequently, the Hamiltonian manifestly depends on time. As shown in the previous section, taking its time derivative explicitly shows that it is not a constant of motion. 

Second, we assume that the variation is produced by an infinitesimal spatial displacement $\boldsymbol{\eta}$ (see Fig.\,\ref{fig:Variation}(b)). Then, if the system is invariant under spatial translations we must have $A'_{p}\left(\mathbf{r},t\right)=A{}_{p}\left(\mathbf{r}-\boldsymbol{\eta},t\right)\simeq A_{p}\left(\mathbf{r},t\right)-\boldsymbol{\eta}\cdot\nabla A_{p}\left(\mathbf{r},t\right)$, so that $dA_{p}\left(\mathbf{r},t\right)=-\boldsymbol{\eta}\cdot\nabla A_{p}\left(\mathbf{r},t\right)$ . Again, substituting this result into (\ref{eq:dS_final}) and factoring out $\boldsymbol{\eta}$
we find that the conserved quantity must be
\begin{equation}
\mathbf{P}=\sum_{p}\int d^{3}r\,\,\frac{\partial\mathcal{L}}{\partial\dot{A}_{p}\left(\mathbf{r},t\right)}\,\nabla A_{p}\left(\mathbf{r},t\right)=-\sum_{p}\int d^{3}r\,\,\varepsilon\left(t\right)\dot{A}_{p}\left(\mathbf{r},t\right)\,\nabla A_{p}\left(\mathbf{r},t\right)
\label{eq:P}
\end{equation}

\noindent which equals the Minkowski momentum in (\ref{eq:P_M_A}). Therefore, we finally found that the fact that time-varying media are invariant under spatial translations directly enforces that the Minkowski momentum is a conserved quantity.

\section{Concluding remarks}

Symmetries play a fundamental role in physics. They reduce the complexity of difficult problems, as well as the computational cost needed to solve them. Symmetries also enable the identification of conserved quantities and the formal link between both symmetries and conserved quantities is Noether's theorem. One of the reasons why time-varying media and/or temporal metamaterials provide a fresh view on electromagnetic theory is because they break temporal symmetries, which are conserved in most traditional photonics systems, while they maintain spatial symmetries. However, the connection between spatial and temporal symmetries and the properties of time-varying media is not always explicitely stated, or analyzed through the point of view of Lagrangian mechanics. The present tutorial aims at filling this gap. Furthermore, we hope that this tutorial may clarify the subtleties of the conservation of the electromagnetic momentum in time-varying media, the nuances of defining the momentum of the electromagnetic fields within the Abraham-Minkowski debate, and that it will foster further research on the role and significance of symmetries in temporal metamaterials.

\bibliography{library}

\end{document}